\documentclass[aps,pra,superscriptaHHress,twocolumn]{revtex4-1}
\usepackage{dcolumn}
\usepackage{amsmath}
\usepackage{float}
\usepackage{amsfonts}
\usepackage{amssymb}
\usepackage{graphicx}
\usepackage{hyperref}
\usepackage{xcolor}
\usepackage{bm} % para bold em símbolos matemáticos
\usepackage{siunitx} % para unidades e grandezas físicas
\usepackage{lipsum}

\begin{document}
\title{Pseudovortex-assisted generation of first-order vector modes}

\author{L. Marques Fagundes}
\affiliation{Departamento de F\'{i}sica, Universidade Federal de Santa Catarina, CEP 88040-900, Florian\'{o}plis, SC, Brazil}
\author{P. Schneider Lacerda}
\affiliation{Departamento de F\'{i}sica, Universidade Federal de Santa Catarina, CEP 88040-900, Florian\'{o}plis, SC, Brazil}
\author{T. A. Petry Kock}
\affiliation{Departamento de F\'{i}sica, Universidade Federal de Santa Catarina, CEP 88040-900, Florian\'{o}plis, SC, Brazil}

\author{N. Rubiano da Silva}
\affiliation{Departamento de F\'{i}sica, Universidade Federal de Santa Catarina, CEP 88040-900, Florian\'{o}plis, SC, Brazil}

\author{P. H. Souto Ribeiro}
\affiliation{Departamento de F\'{i}sica, Universidade Federal de Santa Catarina, CEP 88040-900, Florian\'{o}plis, SC, Brazil}

\author{R. Medeiros de Ara\' ujo}
\affiliation{Departamento de F\'{i}sica, Universidade Federal de Santa Catarina, CEP 88040-900, Florian\'{o}plis, SC, Brazil}

\begin{abstract}
%\red{\lipsum[2]}
A practical method of generating first-order vector modes like radially and azimuthally polarized modes is using the so-called vortex waveplates. Shining these plates with Gaussian beams produces parasitic higher-order modes besides the target vector beam. The reason is that there is a mismatch between the incident Gaussian mode and the doughnut-shaped profile required for the vector beam. This imposes a limit on the fidelity of the generated mode with respect to the target vector beam. In this work we demonstrate theoretically and experimentally that this limit can be overcome if the input Gaussian beam is replaced with a pseudovortex beam, which is a doughnut-shaped mode without topological charge. We use a spatial light modulator to generate this input mode and a geometric-phase-compensated triangular cavity for mode analysis. We demonstrate vector mode fidelity as high as $0.921 \pm 0.013$, which is considerably better than $0.815 \pm 0.003$ obtained with Gaussian input. We also investigate the fidelity as a function of the input beam waist, leading to a theoretical upper bound for the fidelity of $8\pi/27 \approx 0.931$, in close agreement with the experimental results. This approach presents a reliable method of generating high-fidelity vector modes.
%First-order cylindrical vector beams, such as radially and azimuthally polarized modes, are typically generated by illuminating a vortex waveplate with a fundamental Gaussian beam. However, the spatial mismatch between the incident Gaussian profile and the required doughnut-shaped Laguerre-Gauss mode fundamentally limits the modal fidelity and introduces parasitic higher-order transverse modes. In this work, we demonstrate both theoretically and experimentally that the fidelity of the generated first-order vector modes can be significantly enhanced by tailoring the incident field into a pseudovortex beam—a beam possessing a doughnut intensity profile without an azimuthal phase singularity. Using a spatial light modulator and a triangular optical cavity equipped with a geometric phase shifter for mode analysis, we observe the suppression of higher-order transverse resonances, achieving an experimental vector mode fidelity of $0.92 \pm 0.013$, compared to $0.815 \pm 0.003$ obtained with a standard Gaussian input. Furthermore, we investigate the dependence of the modal overlap on the beam waist, confirming that optimal Gaussian illumination yields an upper theoretical bound of $8\pi/27 \approx 0.931$, in close agreement with experimental observations. Our approach provides a versatile and straightforward pathway to synthesize high-purity cylindrical vector beams for optical communication, optical manipulation, and quantum information protocols. 
\end{abstract}

\maketitle

\section{Introduction}

\begin{figure}[h]
    \centering
    \includegraphics[width=\columnwidth]{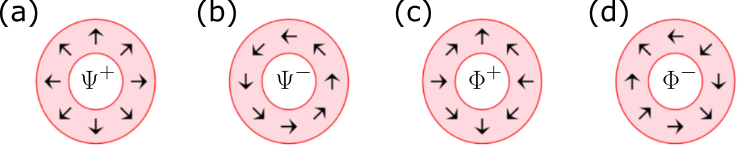}
    \caption{Typical transverse profiles of first-order cylindrical vector beams: (a) radially polarized, (b) azimuthally polarized, and (c-d) hyperbolically polarized. All profiles exhibit a doughnut-shaped intensity distribution with zero intensity at the optical axis.}
    \label{fig:rad-azim}
\end{figure}

In contemporary optics, vector beams are paraxial light fields whose state of polarization varies across the transverse profile \cite{Kumar2014, Rosales2018, Zhan2009}. Among the most widely studied examples are radially and azimuthally polarized cylindrical vector beams (CVB), shown in Fig.~\ref{fig:rad-azim}a and Fig.~\ref{fig:rad-azim}b. Electromagnetic fields exhibiting such polarization structures were already been discussed in the 1950s–1970s, specially in the context of microwave engineering, where radially and azimuthally polarized modes naturally arise as TM${01}$ and TE${01}$ solutions of circular waveguides and antenna systems  \cite{Collin1966}. Closely related field configurations also appeared in early studies of tightly focused electromagnetic fields, where the vectorial nature of light was shown to play a fundamental role in image formation and diffraction \cite{Richards1959,Svensson2025}. Interest in spatially inhomogeneous polarization structures was further stimulated by the development of polarization-singularity theory in the 1980s \cite{Nye1983a,Nye1983b}. %Subsequent work by Berry and Dennis extended the geometrical and statistical description of polarization singularities in vector wave fields \cite{Berry2001}.

Since then, advances in beam generation, manipulation, and detection have transformed vector beams into a central topic within structured light \cite{Zhan2009}, driven by their unique focusing properties \cite{Dorn2001,Svensson2025,Quabis2000}, rich topological structure, and applications in optical trapping \cite{Donato2012,Moradi2019}, super-resolution microscopy \cite{Liu2022, Svensson2025}, laser materials processing \cite{Kozawa2020,svensson2026transformation}, and optical communications \cite{Ndagano2018,li2021liquid}.

Among the various applications of vector beams, their use in optical communications is particularly appealing because the radial and azimuthal polarization states can serve as an orthogonal basis that does not require a shared reference frame between the transmitter and receiver, exploiting their rotational symmetry \cite{Milione2015,Stevens2026}. Beyond their potential for information encoding, these polarization structures also offer opportunities for optical mode manipulation and discrimination \cite{Jia2019,Milione2015,Zhou2024}. In particular, our research group has recently shown that optical cavities provide a promising platform for analysing vector beams, as their mode-dependent resonance properties can be exploited to distinguish between radial and azimuthal polarization states and to separate their superpositions into distinct output channels \cite{Rodrigues2024,Lacerda2025}.

Several approaches have been developed to generate vector beams, including the coherent superposition of orthogonally polarized modes using interferometric arrangements \cite{Maurer2007,Roman2024}, and the use of spatial light modulators (SLMs) to encode spatially varying phase and polarization distributions \cite{Chen2011,Zhang2017}. Other approaches rely on polarization-dependent elements, such as q-plates and patterned liquid-crystal devices, which exploit spatially varying birefringence to couple polarization and orbital angular momentum \cite{Cardano2012,Milione2015,Lim2008, Graffitti2020}. 

A particularly attractive alternative is the use of a vortex waveplate (VWP), also referred to as an S-waveplate or a geometric-phase retarder, which incorporates a spatially varying optical-axis orientation into a single optical element. When operated as a half-wave retarder, such a device can convert a uniformly linearly polarized input beam into a radially or azimuthally polarized beam, depending on the orientation of the input polarization. This single-element approach offers a compact and mechanically stable alternative to more elaborate arrangements, avoiding the need for interferometric recombination and reducing the number of alignment-sensitive components.

In this work, we discuss both theoretically and experimentally how to improve the fidelity of first-order cylindrical vector beams generated by a VWP by correctly choosing the spatial profile of the impinging beam. We call the optimal spatial profile \textit{pseudovortex} \cite{Fagundes2025}, as discussed in the next section. In section \ref{sec:exp}, we present the experimental setup used to generate and analyze the vector beams, which includes vector mode tomography and the analysis of the resonant peaks of a triangular optical cavity. Finally, in section \ref{sec:results} we discuss the results, showing that the use of pseudovortices indeed increase the fidelity of the generated vector beam.

\section{Pseudovortices and Vector Modes}
\label{sec:pseudo}

A convenient basis for first-order cylindrical vector beams (CVB) is obtained by combining the first-order Laguerre-Gauss modes $LG_{\pm 10}$, which have radial index 0 and azimuthal phase factors $\mathrm{e}^{\mp i\varphi}$, with circular polarization states:
\begin{align}
    \mathbf{\Psi}^+&=LG_{10}\,\mathbf{\hat{e}}_R\,+\,LG_{-10}\,\mathbf{\hat{e}}_L\,, \label{eq:Psi+}\\
    \mathbf{\Psi}^-&=LG_{10}\,\mathbf{\hat{e}}_R\,-\,LG_{-10}\,\mathbf{\hat{e}}_L\,, \label{eq:Psi-}\\
    \mathbf{\Phi}^+&=-\left(LG_{-10}\,\mathbf{\hat{e}}_R\,+\,LG_{10}\,\mathbf{\hat{e}}_L\right)\,, \label{eq:Phi+}\\
    \mathbf{\Phi}^-&=-\left(LG_{-10}\,\mathbf{\hat{e}}_R\,-\,LG_{10}\,\mathbf{\hat{e}}_L\right)\,, \label{eq:Phi-}
\end{align}
where $\mathbf{\hat{e}}_R$ and $\mathbf{\hat{e}}_L$ denote the unit vectors associated with right- and left-handed circular polarization, respectively.

All four basis states share the characteristic doughnut-shaped intensity distribution of the $LG_{10}$ mode. Their distinguishing feature lies in the spatially varying polarization, which depends exclusively on the azimuthal coordinate $\varphi$%, measured here from the vertical axis
. The corresponding polarization patterns of all first-order modes $\mathbf{\Psi}^+$ (radial), $\mathbf{\Psi}^-$ (azimuthal), and $\mathbf{\Phi}^+$/$\mathbf{\Phi}^-$ (hyperbolic) are shown in Fig.~\ref{fig:rad-azim}.

The generation of radially and azimuthally polarized beams can be readily achieved by illuminating a vortex waveplate (VWP) with a linearly polarized Gaussian beam. The vortex waveplate introduces an azimuthally varying phase delay between the orthogonal polarization components, thereby coupling the polarization and spatial degrees of freedom and producing the desired vector vortex state. In this configuration, radially and azimuthally polarized beams are obtained directly, with the orientation of the input linear  polarization determining which of the two polarization patterns is generated. The hyperbolically polarized beams can be obtained by introducing an additional half-wave plate (HWP) after the vortex wave plate, with its optical axis appropriately oriented to transform the polarization structure into the corresponding state.

Importantly, obtaining the correct polarization pattern alone is not sufficient to generate the modes $\mathbf{\Psi}^+$, $\mathbf{\Psi}^-$, $\mathbf{\Phi}^+$, and $\mathbf{\Phi}^-$. The radial intensity profile must also be consistent with that of the constituent $LG_{\pm10}$ modes. For a given beam waist parameter $w_0$, the modes $LG_{\pm10}$ and all the CVBs shown in Fig.~\ref{fig:rad-azim} have the same doughnut-shaped intensity profile. We therefore define a pseudovortex as a beam that has the same intensity profile as a vortex beam, such as $LG_{\pm10}$, but does not possess its azimuthal phase profile. Instead, the pseudovortex phase profile is flat. In the present case, pseudovortices provide the appropriate input intensity distribution while allowing the vortex wave plate to generate the required spatially varying polarization. Thus, using pseudovortices rather than zero-order Gaussian beams is essential for reproducing simultaneously the correct intensity distribution and polarization structure of the first-order CVB basis states defined in Eqs.~(\ref{eq:Psi+})--(\ref{eq:Phi-}).

\section{Experimental setup}
\label{sec:exp}

\begin{figure}[h]
    \centering
    \includegraphics[width=\columnwidth]{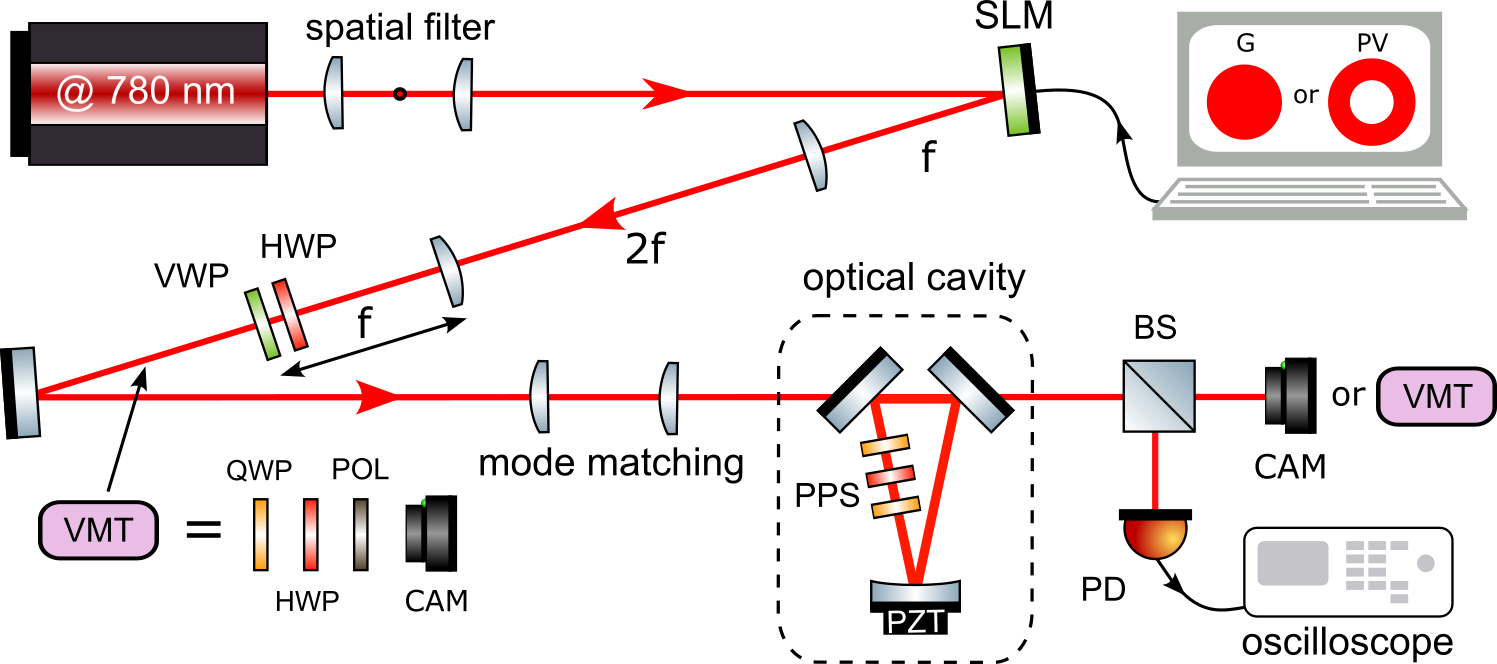}
    \caption{Experimental setup. An extended-cavity diode laser operating at 780 nm is spatially filtered and directed onto an SLM. The SLM generates either a Gaussian or a pseudovortex (G or PV) intensity profile, which is imaged onto a vortex waveplate to produce a vector beam. The resulting vector mode is coupled into a triangular optical cavity incorporating a Pancharatnam phase shifter (PPS). Resonance peaks are monitored with an oscilloscope, while the transmitted mode profiles are recorded with a CMOS camera (CAM). VMT: vector mode tomography; HWP, QWP, and VWP: Half, Quarter, and Vortex waveplates; POL: linear polarizer; BS: beam splitter; PD: bulk photodetector.}
    \label{fig:setup}
\end{figure}

The experimental scheme (Fig.~\ref{fig:setup}) described in this section was designed to compare the mode composition of radially polarized beams generated by a vortex waveplate under two different inputs: a Gaussian beam and a pseudovortex beam.

The input beam is generated with the aid of a spatial light modulator (SLM). An expanded, collimated laser beam is directed onto the SLM, where an amplitude modulation technique based on a variable-depth blazed grating is employed to diffract unwanted optical power away from the regions of interest \cite{Davis1999} and control the beam waist. The encoded masks approximate the transverse profiles of the desired beams, generating either a zero-order Gaussian (G) beam or a pseudovortex (PV) beam, both colimated with a waist $w_0=1.5$~mm, initially. The SLM plane is then imaged onto the vortex waveplate using two lenses of focal length $f$ arranged in a $f-2f-f$ configuration.

The imaging system is particularly important for the pseudovortex beam. Since its transverse profile evolves during propagation, imaging the SLM plane onto the vortex waveplate ensures that the beam reaches the waveplate with a doughnut-shaped intensity profile, thereby generating a beam that is as close as possible to a pure first-order vector mode.

A half-wave plate (HWP) placed before the vortex waveplate (Thorlabs, WPV10L) controls the polarization orientation of the input beam, thereby determining whether the vortex waveplate generates a radially polarized beam, an azimuthally polarized beam, or an arbitrary superposition of the two.

%IN CASE WE NEED TO JUSTIFY EXP IMPERFECTIONS, WE MAY CITE THE MISMATCH BETWEEN THE VPW AND LASER WAVELENGTHS
%The fidelity of the generated vector mode depends critically on the retardance of the waveplate: deviations from the ideal half-wave condition introduce unwanted polarization components and residual ellipticity, thereby reducing the purity of the desired polarization distribution. Improving the fidelity of vector-beam generation therefore requires careful control of the waveplate retardance at the operating wavelength, as well as characterization of the polarization state across the transverse beam profile. These considerations are particularly relevant when vector beams are used as inputs to optical mode discrimination schemes, for which deviations from the ideal polarization structure may affect the measured discrimination performance.

The generated vector beam is then mode-matched into a triangular optical cavity of resonant waist $w_\text{cav}=0.13$~mm, which analyzes and compares the modal composition of the beams. The cavity contains a set of three waveplates arranged as QWP($\pi/4$) + HWP($\theta$) + QWP($\pi/4$). This sequence implements a closed-path transformation on the Poincaré sphere for both horizontal and vertical polarization components, imparting geometric phases of $2\theta$ and $-2\theta$ steradians, respectively. These adjustable phases compensate for the birefringence introduced by the cavity mirrors, as described in Ref.~\cite{Rodrigues2024}.

The cavity length is scanned at 10~Hz by a piezoelectric actuator attached to the concave, high-reflectivity mirror. The transmitted light is collected at one of the plane mirrors. During the scan, the resonance peaks are monitored with an oscilloscope. Once a resonance is identified, the piezo voltage is manually adjusted to maintain the cavity on resonance while the corresponding transverse mode profile is recorded with a CMOS camera (Thorlabs CS165MU).

\section{Results and Discussion}
\label{sec:results}

\begin{figure*}
    \centering
    \includegraphics[width=\textwidth]{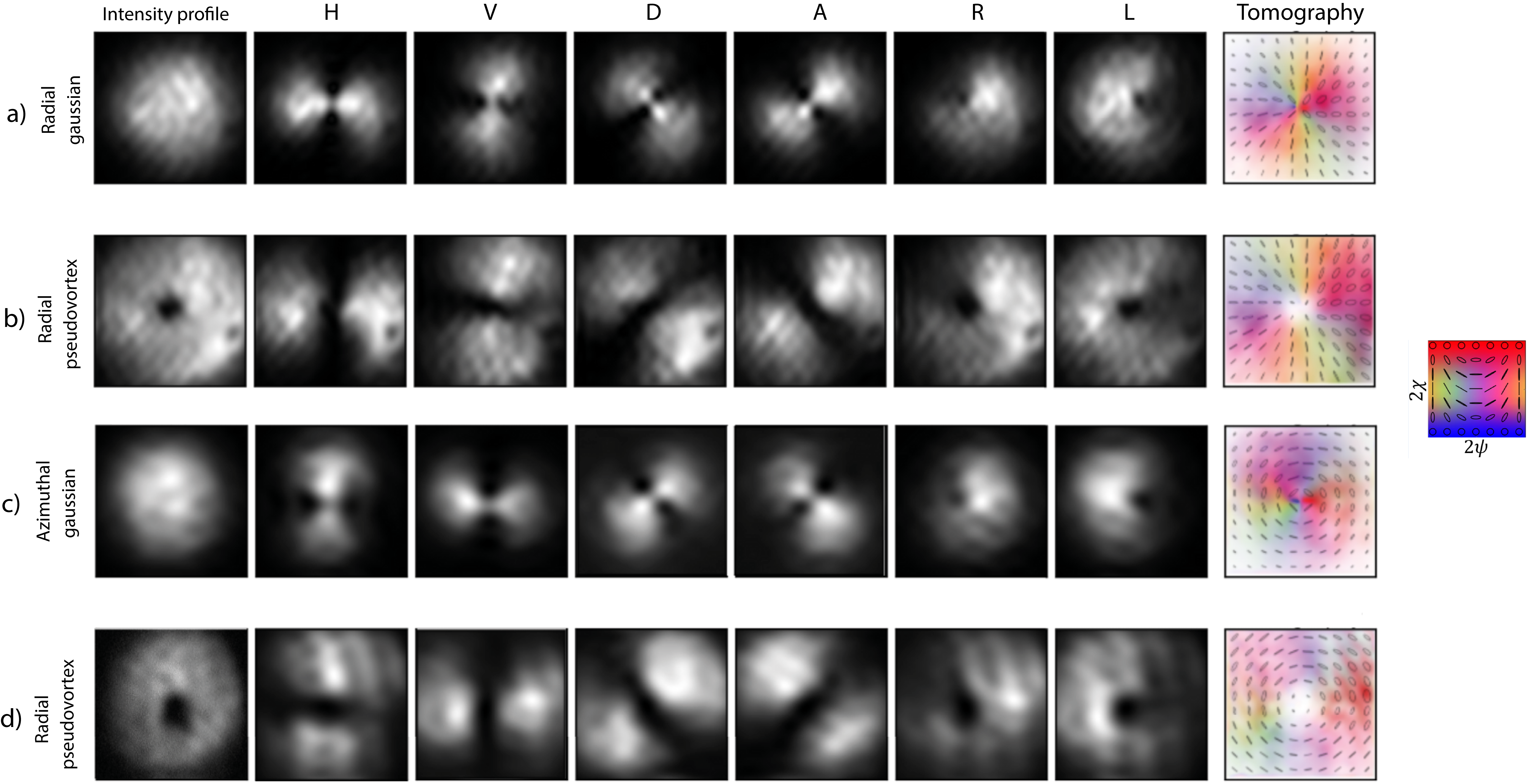}
    \caption{Vector mode tomography (VMT) imediatelly after the Vortex Waveplate (VWP). H, V, D, A, R, and L are the polarization projections of the beams shown in the first column. The last column shows the reconstruction of the vector beam profile from its measured projections. The color scheme used to visualize the polarization state is illustrated in the legend on the right, where $\chi$ and $\psi$ denote the ellipticity and orientation angles on the Poincaré sphere, respectively, while the intensity distribution is encoded through opacity. (a) and (b): Radially polarized beam obtained from the Gaussian and pseudovortex inputs, respectively. (c) and (d): Azimuthally polarized beam obtained from the Gaussian and pseudovortex inputs, respectively.}
    \label{fig:polarimetria}
\end{figure*}

Our goal is to achieve the generation of an ideal first-order radial (or azimuthal) vortex beam $\mathbf{\Psi}^+$ (or $\mathbf{\Psi}^-$). In this section, we compare the vector beams generated by the vortex waveplate under illumination with either a Gaussian beam or a pseudovortex beam. We validate the results using two complementary methods:  performing polarization tomography and analyzing the resonance peaks of the triangular cavity. 

Let us start by analyzing the results of the first method. Fig. \ref{fig:polarimetria} shows the results of the polarization tomography of the beam immediately after the VWP. The images were recorded with a CMOS camera and filtered to suppress Fourier components associated with interference fringes arising from reflections at the camera surfaces. A lens was used to image the plane of the VWP to the plane of the camera. 

The first column of Fig. \ref{fig:polarimetria} shows images of the overall intensity of the radially/azimuthally polarized beams generated with Gaussian and pseudovortex beams. As expected, the pseudovortex delivers a beam with zero intensity at its center, approximating the doughnut shape of a first-order radial beam $\mathbf{\Psi}^+$, whereas the Gaussian beam fails to do the same. The following columns display the projections of the resulting beams onto six polarizations: H (linear at $0^\circ$), V (linear at $90^\circ$), D (linear at $45^\circ$), A (linear at $-45^\circ$), R (circular right) and L (circular left). Notice, in particular, that the linear polarization projections (H, V, D, A) applied upon the beams generated by a pseudovortex (Fig. \ref{fig:polarimetria}b and \ref{fig:polarimetria}d) result in the typical well separated lobes of Hermite-Gaussian beams, as expected. In contrast, in the gaussian case (Fig. \ref{fig:polarimetria}a and \ref{fig:polarimetria}c), these projections show lobe-like structures connected at the optical axis. The last column, called tomography, displays the polarization profiles, along with intensity, reconstructed from the projections in each line. The tomography consistently reproduces the intensity profiles of the first column and correctly captures the intended polarization structure (radial or azimuthal). Slightly elliptical polarization states arise locally due to imperfections in the optical elements, without affecting the overall results.

\begin{figure}
    \centering
    \includegraphics[width=0.9\columnwidth]{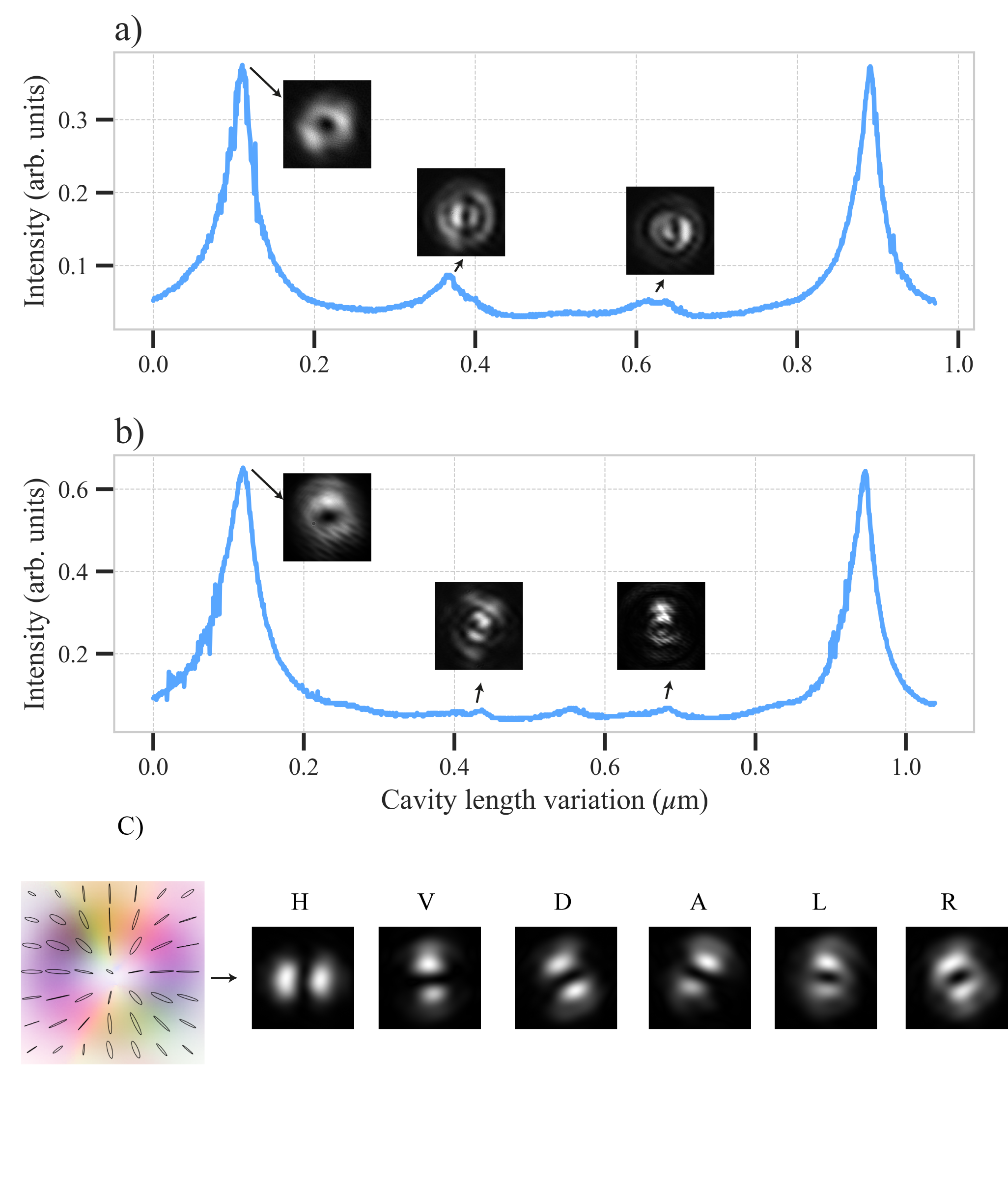}
    \caption{Radially polarized beam resonances in a triangular cavity. Comparison of the resonance peaks when the input beam to vortex waveplate is (a) a zero-order Guassian beam and (b) a pseudovortex beam; (c) displays the vector mode tomography of the resonant $\mathbf{\Psi}^+$ beam.}
    \label{fig:picos+}
\end{figure}

\begin{figure}
    \centering
    \includegraphics[width=0.9\columnwidth]{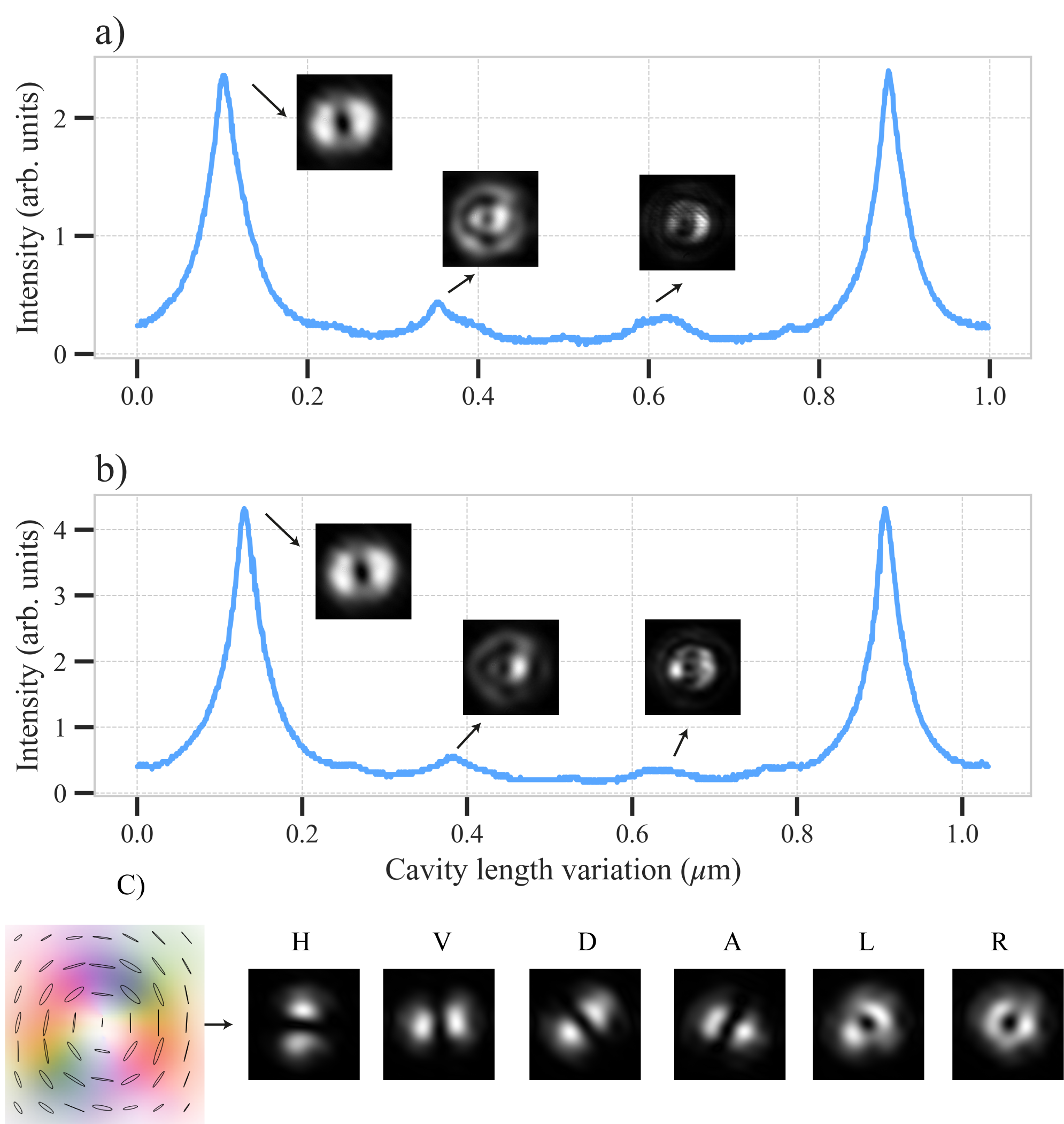}
    \caption{Azimuthally polarized beam resonances in a triangular cavity. Comparison of the  resonance peaks when the input beam to vortex waveplate is (a) a zero-order Guassian beam and (b) a pseudovortex beam; (c) displays the vector mode tomography of the resonant $\mathbf{\Psi}^-$ beam.}
    \label{fig:picos-}
\end{figure}

The second method consists of sending the beams generated by the VWP to a triangular optical cavity and analyzing \textit{i}) the resonance peaks with a bulk detector and \textit{ii}) the corresponding spatial modes observed in the camera. From the resonance peaks, we are able to assess the fidelity $F$ of the beam generated by the VWP. The fidelity of a beam with a transverse field $\mathbf{E}$ with respect to the first-order radial beam (transverse field $\mathbf{\Psi}^\pm$) is mathematically defined as the squared modulus of the overlap integral of the two modes:
\begin{equation}
    F^\pm =\left|\int_{\mathbb{R}^2}(\mathbf{\Psi}^\pm(x,y))^*\cdot \mathbf{E}(x,y) \ dx\,dy\right|^2,
    \label{eq:fidelity}
\end{equation}
where the symbol $\cdot$ represents the scalar product between the vector fields. The fidelity can assume values from 0 to 1 and measures the fraction of energy in the field $\mathbf{E}$ that comes from the mode $\mathbf{\Psi}^\pm$. In a triangular cavity, if a given peak corresponds exclusively to the mode $\mathbf{\Psi}^+$ (or $\mathbf{\Psi}^-$), then identifying this peak, measuring its height and comparing it to the sum of the heights of all measured peaks should give access to the value of the fidelity. In fact, that is how we measure the fidelity experimentally:
\begin{equation}
    F^\pm_\text{exp} =\frac{\text{height of }(\mathbf{\Psi}^\pm)\text{'s peak}}{\Sigma \text{ heights of all peaks}}\,.
    \label{eq:fidelity-exp}
\end{equation}

Figure \ref{fig:picos+}(a) shows the resonance peaks obtained when a radially polarized vector beam generated from a Gaussian input beam is sent through the triangular optical cavity. The beam exhibits three main resonance peaks. The tallest peak on the left corresponds to the first-order radial vector mode $\mathbf{\Psi}^+$, whereas the two smaller peaks on its right arise from measurable contributions of third- and fifth-order modes, respectively. The tall peak on the right is a repetition of the leftmost peak, reflecting the periodicity of the resonance spectrum with a period of $\lambda$. Using these data and Eq. (\ref{eq:fidelity-exp}), we obtain $F^+_\text{exp}=0.815 \pm 0.003$, which is comparable to the theoretical prediction $F^+=0.785$ from Eq. (\ref{eq:fidelity}). When the Gaussian input beam is replaced with a pseudovortex beam in the VWP (Fig. \ref{fig:picos+}(b)), the higher-order peaks are substantially suppressed, indicating a reduction in the contributions of higher-order modes. In this case, the theoretical prediction yields the maximum possible fidelity, $F^+=1$. Experimentally, we obtain $F^+_\text{exp}=0.921 \pm 0.013$. Finally, the vector-mode tomography shown in Fig. \ref{fig:picos+}(c) confirms that the dominant resonance peaks correspond to the $\mathbf{\Psi}^+$ vector mode, ruling out alternative scalar or vector modes with the same intensity profile. Figure \ref{fig:picos-} presents analogous results for the generation of an azimuthally polarized beam. The so-obtained fidelities $F^-_\text{exp}$ are $0.809 \pm 0.006$ and $0.919 \pm 0.004$ for the Gaussian and pseudovortex inputs, respectively. These results show that, for both radial and azimuthal beams, the use of a pseudovortex increases the fidelity from around 0.8 to about 0.9.

It is important to recall that the zero-order Gaussian beam and the pseudovortex beam used have both the same waist size $w_0=1.5$~mm on the VWP. Between the VWP and the optical cavity, there is a pair of lenses performing the mode matching that brings the beams to a waist size $w_{cav}=0.13$~mm, which is the cavity resonant beam waist. We know that this particular value of $w_0$ optimizes cavity coupling for a linearly polarized zero-order gaussian beam. 

A legitimate question that arises is whether this value $w_0=1.5$~mm also optimizes the coupling of a radially/azimuthally polarized gaussian beam to the triangular cavity. Put another way, can we increase the fidelity to a first-order mode $\mathbf{\Psi}^\pm$ by changing the waist size of the input Gaussian beam? In the following, we address this question for the case of radial polarization both theoretically and experimentally. 

Let $w_g$ be the waist of a Gaussian beam impinging on the VWP and $w_0$ be the desired waist for the vector mode $\mathbf{\Psi}^+$. The calculated fidelity as a function of $w_g$ (see Appendix \ref{app:overlap}) is
\begin{equation}
    F^+ = \frac{2\pi w_0^2w_g^4}{\left(w_0^2+w_g^2\right)^3}\,,
    \label{eq:fidelity-wg}
\end{equation}
which reaches its maximum of $8\pi/27\approx 0.931$ at $w_g=\sqrt{2}w_0$.

\begin{figure}
    \centering
    \includegraphics[width=1\linewidth]{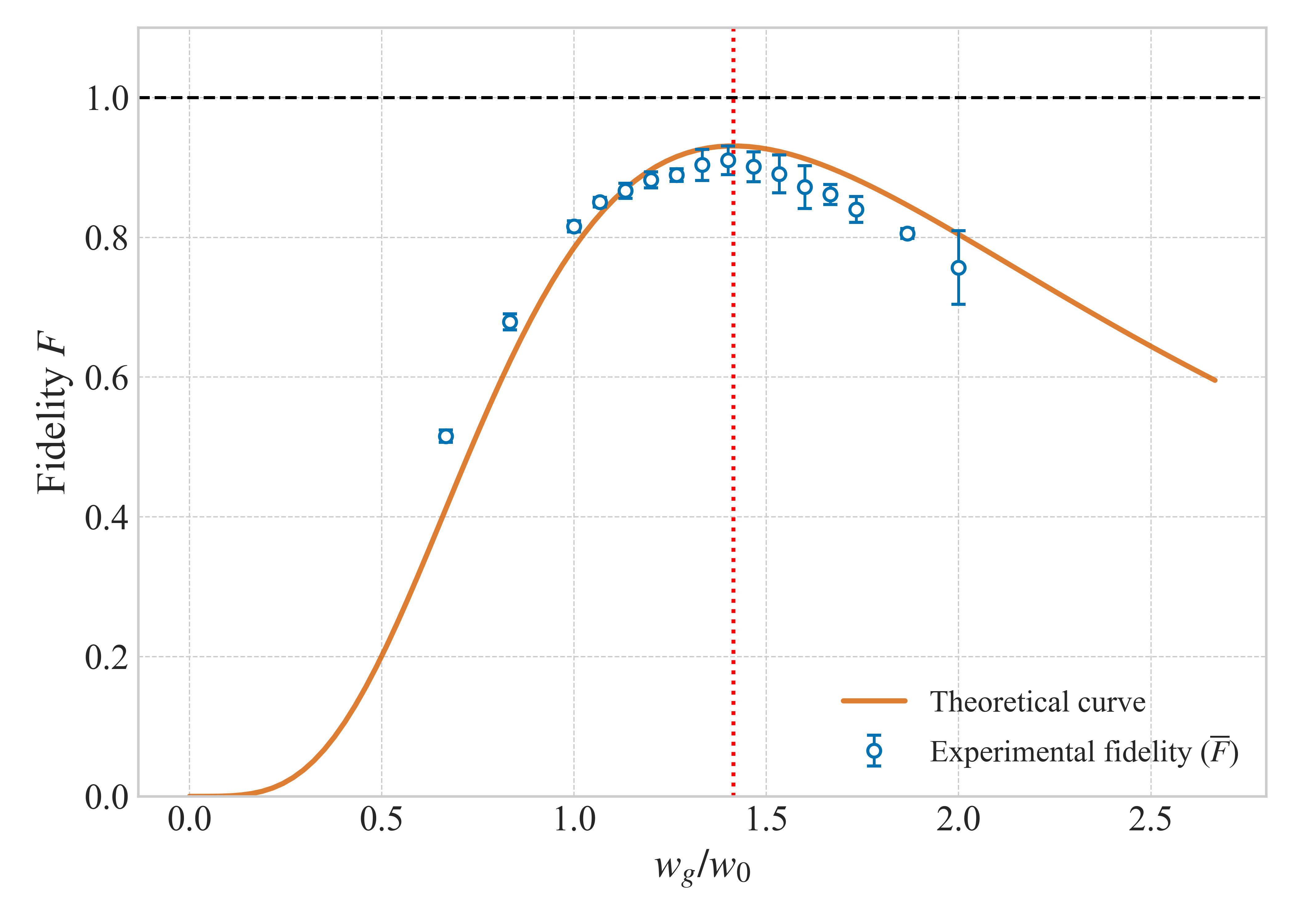}
    \caption{Fidelity of the radial beam generated by Gaussian beams with different waist sizes. The blue curve shows the theoretical fidelity obtained using equation (\ref{eq:fidelity-wg}). The experimental points show the average fidelity for that waist over five measurements, with the error bars being the standard deviation of the measurements. The red dashed line represents $w_g/w_0=\sqrt{2}$, the value which optimizes the fidelity. The black dashed line represents the maximum fidelity possible, $F=1$.}
    \label{fig:overlap}
\end{figure}

This theoretical result was tested experimentally as follows. Using the SLM, we varied the Gaussian beam waist, $w_g$, from approximately $w_0/2$ to $2w_0$. For each value of $w_g$, we measured the experimental fidelity $F^+_\text{exp}$ according to Eq. (\ref{eq:fidelity-exp}). Figure \ref{fig:overlap} shows both the theoretical and experimental fidelities as a function of $w_g$. The results show close agreement between theory and experiment, with both reaching a maximum around $w_g=\sqrt{2}w_0$. The maximum measured fidelity was $F^+_\text{exp} \text{(max)} = 0.910\pm 0.009$, which is numerically lower, but within experimental uncertainty, than the fidelity obtained with a pseudovortex.

\section{Conclusion}
%\red{\lipsum[2]}
In this work, we showed that the mode fidelity of first-order cylindrical vector beams produced by vortex waveplates depends strongly on the transverse shape of the input beam. Although a standard Gaussian beam provides the required inhomogeneous polarization pattern, it lacks the central intensity null of the target vector mode. This spatial mismatch inevitably couples energy into higher-order transverse modes, limiting the achievable fidelity.

We demonstrated that this problem can be mitigated by shaping the incident field into a pseudovortex beam, which reproduces the doughnut intensity profile without introducing any phase singularity. By using a triangular optical cavity as a modal analyser, we observed a clear suppression of unwanted transverse resonances for a radially polarized beam and measured an experimental fidelity of $0.921 \pm 0.013$, compared to $0.815 \pm 0.003$ obtained with the standard Gaussian input. Very similar results were obtained in the generation of an azimuthally polarized beam. Furthermore, our study of the fidelity as a function of the Gaussian waist confirmed the theoretical maximum of $8\pi/27 \approx 0.931$ around $w_g = \sqrt{2}w_0$, showing a very good agreement between theory and experiment. These results demonstrate that spatial amplitude shaping offers a practical and robust way to generate high-purity vector beams for structured light applications. 

\begin{acknowledgements}
This work has been supported by the following Brazilian research agencies: Conselho Nacional de Desenvolvimento Cient\'{\i}fico e Tecnol\'ogico (CNPq - DOI 501100003593), Coordena\c c\~{a}o de Aperfei\c coamento de Pessoal de N\'\i vel Superior (CAPES DOI 501100002322), Funda\c c\~{a}o de Amparo \`{a} Pesquisa do Estado de Santa Catarina (FAPESC - DOI 501100005667, DOI 2025TR001683), Instituto Nacional de Ci\^encia e Tecnologia de Informa\c c\~ao Qu\^antica (INCT-IQ 465469/2014-0, INCT-IQNano 406636/2022-2 and INCT-DQ 408783/2024-9).  
\end{acknowledgements}

\appendix
\section{Gaussian-Radial overlap calculation}
\label{app:overlap}

To calculate the overlap between the Gaussian-genereted radial beam $\mathbf{E}_g$ and the first-order radial beam $\mathbf{\Psi}^+$, we begin by writing their normalized vectorial fields at the beam waist ($z=0$):
\begin{align}
    \mathbf{E}_g &= \sqrt{\frac{2}{\pi w^2_g}}\left(\frac{x\,\mathbf{\hat{x}}+y\,\mathbf{\hat{y}}}{\sqrt{x^2+y^2}}\right)e^{-\frac{x^2+y^2}{w_g^2}}\,,\\
     \mathbf{\Psi}^+ &= \frac{2}{\sqrt{\pi}w_0^2}(x\,\mathbf{\hat{x}}+y\,\mathbf{\hat{y}})e^{-\frac{x^2+y^2}{w_0^2}}\,.
\end{align}

The overlap integral is given by the inner product of the two vector fields over the transverse plane, $I = \int_{\mathbb{R}^2} (\mathbf{\Psi}^+)^* \cdot \mathbf{E}_g \,dx\,dy$, which yields:
\begin{equation}
    I = \frac{2\sqrt{2}}{\pi w_0^2 w_g}\iint_{-\infty}^{+\infty}\sqrt{x^2+y^2}e^{-(x^2+y^2)\left(\frac{1}{w_0^2}+\frac{1}{w_g^2}\right)}dxdy.
\end{equation}

By changing to polar coordinates, the integral yields:
\begin{equation}
    I = \frac{\sqrt{2\pi}w_0w_g^2}{\left(w_0^2+w_g^2\right)^{3/2}}.
\end{equation}

Differentiating this expression with respect to $w_g$ reveals that the optimal waist for the Gaussian beam is $w_g = \sqrt{2}w_0$. Substituting this condition back into the overlap equation provides the maximum theoretical overlap between a Gaussian-generated radial beam and the $\mathbf{\Psi}^+$ mode:
\begin{equation}
    I_{max} = \frac{2\sqrt{2\pi}}{3\sqrt{3}}\approx0.965\,,
\end{equation}
which corresponds to a maxfidelity (see eq. \ref{eq:fidelity}) of 
\begin{equation}
    F_{max} = \frac{8\pi}{27}\approx0.931\,.
\end{equation}

\newpage
\bibliographystyle{apsrev4-2}
\bibliography{biblio}

 %%%%%%%%%%%%%%%%%%%%%%%%%%%%%%%%%%%%%%%%%%%%%%%%%%%%%%%%%%

\end{document}